\documentclass[pre,twocolumn,superscriptaddress,floatfix]{revtex4-1}  

\usepackage[paperwidth=210mm,paperheight=297mm,centering,hmargin=2cm,vmargin=2.5cm]{geometry}

\usepackage[pdftex]{graphicx}
\usepackage{amsmath,amsfonts,amssymb}
\usepackage{natbib}
\usepackage{MnSymbol}
\usepackage{csquotes}

\usepackage{hyperref}
\usepackage{xcolor} 
\usepackage{array}
\usepackage{pgfplots}
\usepackage{graphicx}
\usepackage[left,modulo,pagewise]{lineno}
\usepackage{dsfont}

\definecolor{bluemoi}{rgb}{0.25,0.50 ,0.75} 

\hypersetup{
	bookmarksopen=true,
	pdftitle=" ",
	pdfauthor=" ", 
	pdfsubject=" ", 
	pdftoolbar=true, 
	pdfmenubar=true, 
	pdfhighlight=/O, 
	colorlinks=true, 
	pdfpagemode=UseNone, 
	pdfpagelayout=SinglePage, 
	pdffitwindow=true, 
	linkcolor=bluemoi, 
	citecolor=bluemoi, 
	urlcolor=bluemoi 
}

\makeatletter
\renewcommand{\figurename}{\sf \textbf{Figure}}
\renewcommand{\thefigure}{\arabic{figure}}
\renewcommand{\fnum@figure}{\sf\textbf{\figurename}~\textbf{\thefigure}}
\renewcommand{\tablename}{\sf\textbf{Table}}
\renewcommand{\thetable}{\arabic{table}}
\renewcommand{\fnum@table}{\sf\textbf{\tablename}~\textbf{\thetable}}
\makeatother

\begin{document}
	
\title{Mapping livestock movements in Sahelian Africa} 

\author{Camille Jahel}
\thanks{Corresponding authors: maxime.lenormand@inrae.fr \& camille.jahel@cirad.fr who contributed equally to this work.}
\affiliation{CIRAD, UMR  TETIS, Montpellier, France}
\affiliation{ISRA,BAME, rue des Hydrocarbure, Dakar, Senegal}
\affiliation{PPZS, Pastoral Systems and Dry Lands, Dakar, Senegal}

\author{Maxime Lenormand}
\thanks{Corresponding authors: maxime.lenormand@inrae.fr \& camille.jahel@cirad.fr who contributed equally to this work.}
\affiliation{TETIS, Univ Montpellier, AgroParisTech, Cirad, CNRS, INRAE, Montpellier, France}

\author{Ismaila Seck}
\affiliation{DSV, Dakar, Senegal}
\affiliation{FAO, Regional Office}

\author{Andrea Apolloni}
\affiliation{CIRAD, UMR  ASTRE, Campus  International de Baillarguet, 34398  Montpellier, France}
\affiliation{ISRA, LNERV, rue fond de terre, Dakar, Senegal}
	
\author{Ibra Toure}
\affiliation{CIRAD, UMR  SELMET, Campus  International de Baillarguet, 34398  Montpellier, France}
	
\author{Coumba Faye}
\affiliation{DSV, Dakar, Senegal}
	
\author{Baba Sall}
\affiliation{DSV, Dakar, Senegal}

\author{Mbargou Lo}
\affiliation{DSV, Dakar, Senegal}

\author{C{\'e}cile Squarzoni Diaw}
\affiliation{CIRAD, UMR  ASTRE, Campus  International de Baillarguet, 34398  Montpellier, France}

\author{Renaud Lancelot}
\affiliation{CIRAD, UMR  ASTRE, Campus  International de Baillarguet, 34398  Montpellier, France}

\author{Caroline Coste}
\affiliation{CIRAD, UMR  ASTRE, Campus  International de Baillarguet, 34398  Montpellier, France}

\begin{abstract}
In the dominant livestock systems of Sahelian countries herds have to move across territories. Their mobility is often a source of conflict with farmers in the areas crossed, and helps spread diseases such as Rift Valley Fever. Knowledge of the routes followed by herds is therefore core to guiding the  implementation of preventive and control measures for transboundary animal diseases, land use planning and conflict management.  However, the lack of quantitative data on livestock movements, together with the high temporal and spatial variability of herd movements, has so far hampered the production of fine resolution maps of animal movements. This paper proposes a general framework for mapping potential paths for livestock movements and  identifying areas of high animal passage potential for  those movements. The method consists in combining the information contained in livestock  mobility networks with landscape connectivity, based on different mobility conductance layers. We illustrate our approach with a livestock mobility network in Senegal and Mauritania in the 2014 dry and wet seasons.
\end{abstract}

\maketitle

\begin{figure*}[!ht]
	\centering
	\includegraphics[width=11cm]{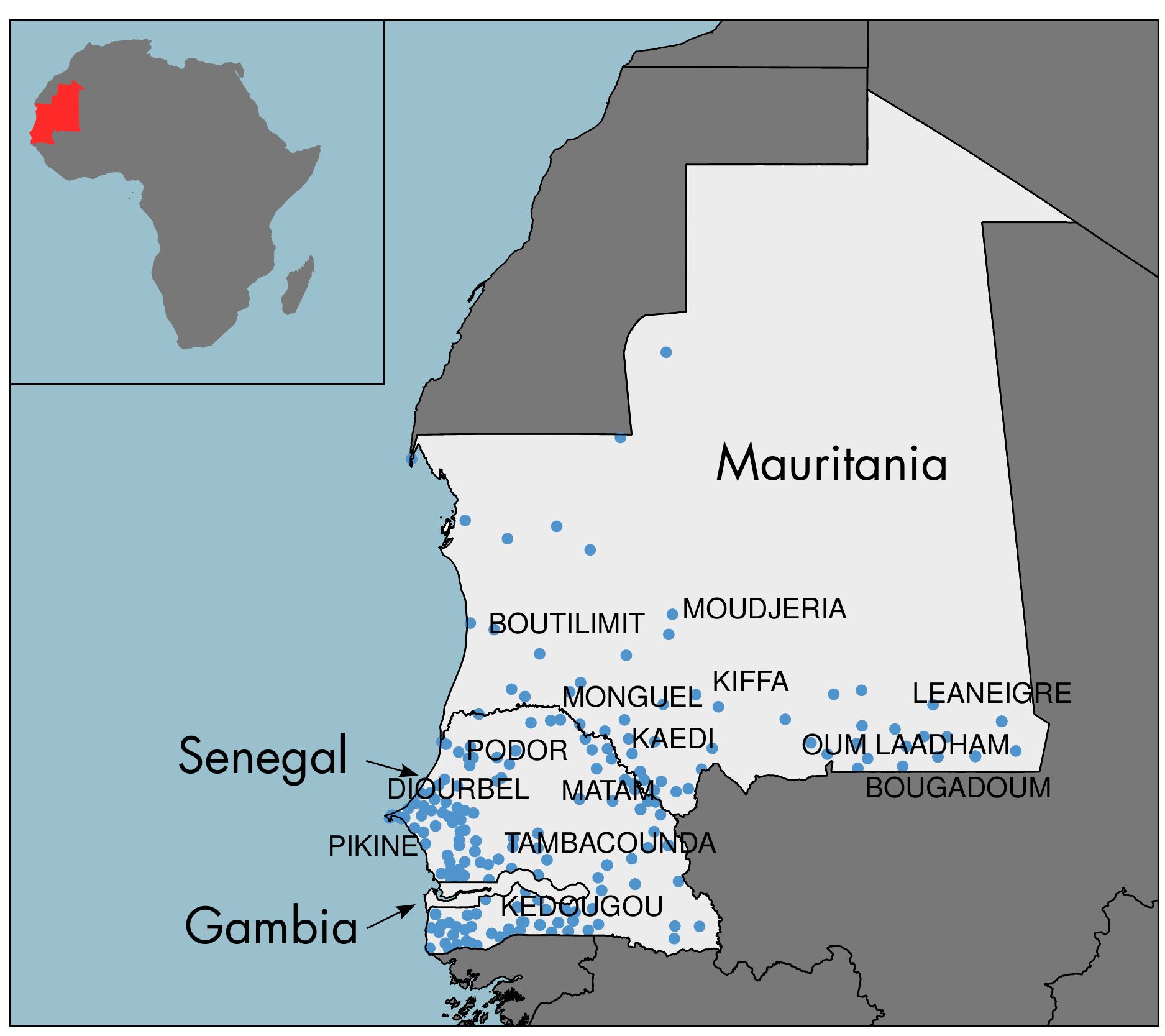} 
	\caption{\textbf{Positions of the nodes of the livestock mobility network.} Each point corresponds to a market represented by a node in the livestock mobility network. The inset shows the location of the studied area in Africa. \label{Fig1}}
\end{figure*}

\section*{Introduction}

Every year in West Africa, millions of animals move from the Sahelian semi-arid regions, where they were bred, towards southern regions looking for better grazing areas, or to be sold on consumption markets \cite{Corniaux2014,Kamuanga2008,Corniaux2016}. These movements often cause conflicts with farmers, especially during the wet growing season, when animals can invade cultivated plots \cite{Moritz2006,Turner2011a,Turner2011b}
. Livestock trade mobility is also a key driver in spreading animal diseases. Indeed, on their way, livestock may cross areas with a high prevalence of mosquitos  (lowlands, wetlands), which are vectors of diseases.  The contact between  animals when herds meet each  other, is also conducive to disease transmission. Mapping movement patterns is thus essential for improving many aspects of livestock management at regional and national level, such as the management of natural resources, the positioning of borehole installations, the reduction of conflicts, and the control of animal diseases. However, the intrinsic complexity of livestock mobility paths makes it extremely tricky to map them. 

One way of mapping livestock spatial distribution consists in working from a census or estimation of the number of animals at different resolutions. Some recent work improved the mapping of static livestock distribution by disaggregating census counts of animals, but provided no information about their actual movements. For instance, Tran et al. \cite{Tran2016} disaggregated census data taken at administrative level to produce risk maps for Rift Valley fever and Napp et al. \cite{Napp2018} used buffer areas to disaggregate their static data. Fourni{\'e} et al. \cite{Fournie2018} used densities derived from human demographic data, aggregated at village level, to study the transmission of  \enquote{Peste des Petits Ruminants}. However, these approaches are limited to a static vision and do not enable animal movements to be explicitly taken into account. 

We recently witnessed the emergence of network-based approaches to study livestock movements \cite{Volkova2010,Bajardi2012,Hardstaff2015}. Such methods have been tested in many African countries \cite{Dean2013,Motta2017,Nicolas2018,Apolloni2018,Mekonnen2019,Chaters2019}. It consists in describing livestock movements as a directed and weighted spatial network, where nodes represent villages, markets or premises and each link between two nodes represents at least one animal moving from one site to another. The weight of a link is equal to the total number of animals exchanged. In some ideal cases, the spatial pathway of the links is known, thanks to GPS tracking of animals \cite{Guo2012}, but in Sahelian areas such data are rarely available and have only been tested on a few cattle \cite{Adriansen2005,Motta2018}. Thus, the majority of livestock network analysis studies do not explicitly spatialize animal pathways between two nodes; the flows of the graph only provide information about the direction, distance and volume of movements. 

Here we propose a way of mapping livestock movements that combines the information contained in livestock mobility networks with a landscape connectivity-based approach. The method consists in producing a conductance map representing the ease of livestock movements, to be linked with the mobility network in order to produce a map of potential paths. We illustrate our approach with a livestock mobility network in Mauritania and Senegal during the 2014 dry and wet seasons. The next section presents the proposed framework and the data used to illustrate our approach. The results are then presented, demonstrating the capacity and robustness of our approach in identifying potential paths for livestock movements in Sahelian Africa. Lastly, we discuss the advantages and limitations of our approach.
	
\section*{Material and methods}

\subsection*{Study area}

Our study area encompasses Senegal and Mauritania, where a recent report estimated the total number of cattle to be between 2 and 3 million \cite{Aubauge2017}. In Mauritania, rangelands are predominant, with agricultural areas being limited to irrigable or flooded areas along the Senegal River and in oases. In Senegal, livestock farming is mostly located in Ferlo, a region of 70,000 km$^2$ in the North east of the country, where climatic conditions do not allow the development of agricultural activity. A large share of the cattle spend the wet season in this rangeland area of Mauritania and northern Senegal, then moves towards the markets, or towards the crop residues of the central and southern regions, especially in the groundnut basin of Senegal. This animal trade mobility network between Mauritania and Senegal involves up to 1.9 million bovines \cite{Apolloni2018}. Fewer than  20\% of these animals are conveyed by vehicles, mostly commercial requests for religious feasts, with the rest traveling on foot, over a distance  of one to three hundred kilometers \cite{Apolloni2018}. Conveyance on foot enables the cattle to benefit from the pastures and crop residues of southern regions in order to continue fattening along the way. Animals traveling on foot often cross large areas before arriving at their final destination. At the border, large cattle herds will cross at official passage points, but the majority of herders use non-official points to avoid paying taxes, or because they are more accessible \cite{Apolloni2018}, increasing the difficulty of mapping their paths.

\begin{figure*}[!ht]
	\centering 
	\includegraphics[width=13cm]{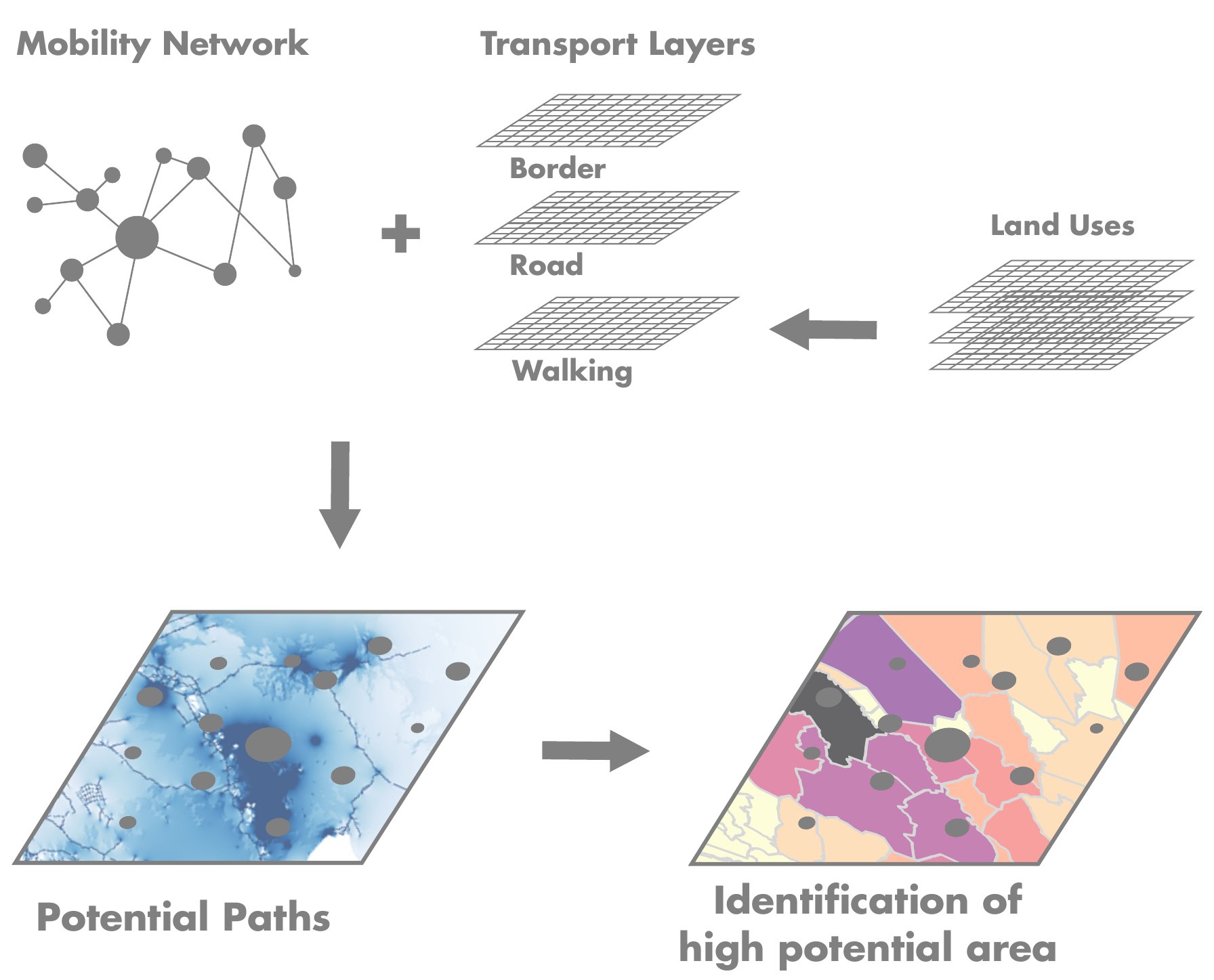}
	\caption{\textbf{Methodology used to map potential paths for livestock movements and identify areas with a high potential for livestock movements based on mobility network and land use information.} \label{Fig2}}
\end{figure*}

\subsection*{Livestock mobility network} 

Livestock mobility data are collected by field Veterinarian Services in Senegal, Gambia and Mauritania. In those countries, a certificate system based on sanitary movement permits (Sanitary \enquote{Laissez-Passer} or LPS) has been set up to keep track of animal mobility and map the main axes of movements in the area. Every time herders move their herds towards markets, or to other grazing area, a certificate is issued declaring, among other things, the date, the location of origin, the location of destination, the species and number of head, and the means of transportation. In this article, we consider only information relative to cattle movements, on foot, in 2014. We aggregated our data on a timescale of one month, providing a representation of the mobility dynamics over the year. This mobility information is represented by a weighted and directed livestock mobility network where the nodes correspond to the origin and destination locations (Figure \ref{Fig1}), and a directed link exists between two nodes if at least one animal is exchanged from one location to another. A link is characterized by the number of head exchanged (volume) and the month of occurrence. We distinguished between the characteristics of the network during the wet season (June to October) and the dry season (November to May).

We used several centrality metrics to analyze the weighted and directed livestock mobility network described above. We focused on five measures, the in- and out-degree (total number of links ingoing to a node or outgoing from a node, respectively), the in- and out-strength (total number of animals ingoing to a node or outgoing from a node, respectively), and the betweenness. The betweenness of a node is proportional to the number of shortest paths (weighted by the distance) going through this node. 

\begin{figure*}[!ht]
	\centering
	\includegraphics[width=14cm]{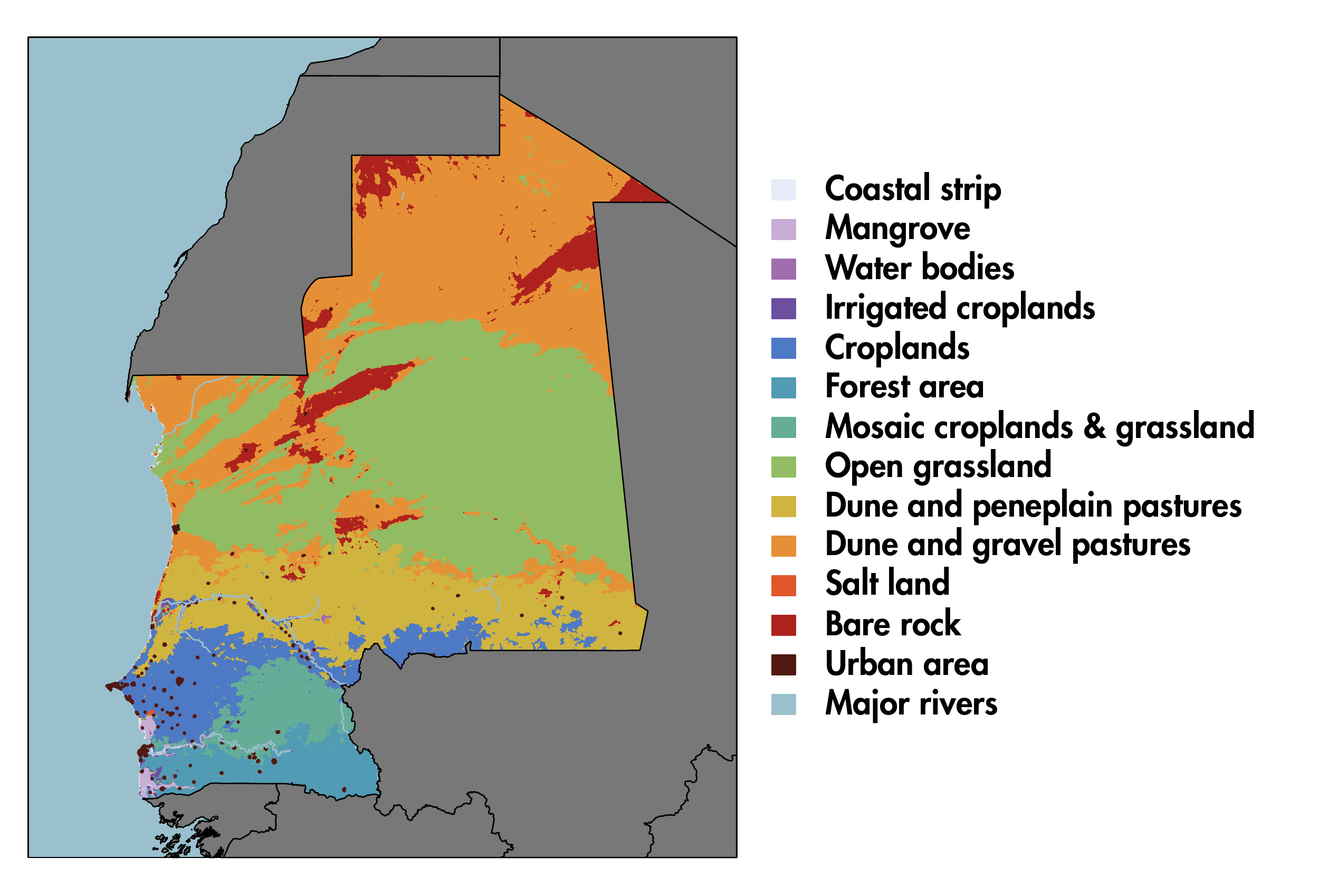} 
	\caption{\textbf{Land use map.} \label{Fig3}}
\end{figure*}

\subsection*{Mapping potential paths for livestock movements} 

As depicted in Figure \ref{Fig2} the main purpose of the proposed methodology is to combine the information contained in the livestock mobility network described above and land use information to map the potential paths for livestock movements at high spatial resolutions. This section describes in detail the methods used to build the conductance map and to assign a potential route between every pair of nodes of our livestock mobility network based on this conductance map. Hereinafter referred to as landscape connectivity approach.\\

\noindent\textbf{Conductance map.} We used land-use/land-cover information and transportation features in Senegal and Mauritania to develop conductance maps represented as rasters at 500-meter resolution. Conductance is the reciprocal of resistance and therefore represents a greater ease of livestock movements. We assigned to each pixel of the conductance map a value according to its livestock movement propensity, ranging from 0 (low conductance/high resistance) to 1 (high conductance/low resistance). It is important to note that a pixel with no value (see Table \ref{Tab1}) means that it is not possible to go through this pixel. We then applied an iterative process based on three different levels of information described below. Each geographical layer was rasterized to the same extent with a pixel dimension of $500 \times 500$ m$^2$.

\begin{itemize}
	\item A walking layer based on land use and land cover information provided by the FAO (data available online at \url{http://www.fao.org}, last accessed 14/06/2019). The original classification has been aggregated in 14 land-use types available in Figure \ref{Fig3} and Table \ref{Tab1}.
	\item The main road network in Senegal, Gambia and Mauritania downloaded from OpenStreetMap (data available online at \url{https://www.openstreetmap.org}, last accessed 18/02/2020). A map of the road network is available in Appendix (Figure S1).
	\item The administrative border line between Senegal and Mauritania comes from the GADM web platform (data available online at \url{https://gadm.org/}, last accessed 18/02/2020). The border crossing points (red points in Figure S1 in Appendix) were given by an expert from a Senegalese's institute specialist in cattle mobility.
\end{itemize}

The bottom level of information regarding livestock movements is called the walking layer $W$. On this layer, conductance is based on landscape features and changes according to the season. We relied on expert knowledge to assign a conductance weight to each type of land use (Table \ref{Tab1}). To do so, we conducted four individual interviews with experts, asking them to rank and then estimate the conductance value of different types of land use according to their knowledge of breeder mobility strategies. We analyzed the results with a fifth expert to choose the final values. The experts were researchers from French or Senegalese institutes and were specialists in cattle mobility, or members of Senegalese governmental institutions in the livestock sector. 

\begin{table}[!h]
	\caption{\textbf{Land use weights according to the season.} The weights represent the conductance from 0 (low conductance/high resistance) to 1 (high conductance/low resistance). The symbol '-' (no value) indicates that no movement is possible.}
	\label{Tab1}
	\begin{center}
		\begin{tabular}{lcc}
			\hline
			Type & Dry season & Wet season \\
			\hline
			Coastal strip & 0.5 & 0.5 \\
			Mangrove & 0.25 & 0.25  \\
			Water bodies & 0.5 & 0.25  \\
			Irrigated croplands& - & -  \\
			Croplands & 1 & 0.125  \\
			Forest area  & 0.5 & 0.5  \\
			Mosaic croplands \& grassland & 1 & 0.5  \\
			Open grassland  & 1 & 1  \\
			Dune and peneplain pastures  & 0.875 & 0.375  \\	
			Dune and gravel pastures  & 1 & 0.875  \\
			Salt land  & 1 & 0.75  \\
			Bare rock & 0.75 & 0.75  \\
			Urban area & 0.125 & 0.125  \\
			Major rivers & - & -  \\
			\hline
		\end{tabular}
	\end{center}
\end{table}

The second level of information is represented by the main road network in Senegal and Mauritania. It is combined with the walking layer assigning the conductance value 1 (high conductance/low resistance) to any pixels of $W$ crossed by a road to obtain a new layer $R$. Note that the influence of $W$ on $R$ can be adjusted with the parameter $\delta_W \in [0,1]$. More formally, the value $R_i$ of a pixel $i$ according to the walking layer $W$ and $\delta_W$, is defined as follows,
\begin{equation}
\label{R}
R_i =
\begin{cases}
1 & \text{if a road cross } i\\
\delta_W  W_i & \text{otherwise}
\end{cases}
\end{equation}
Finally, the last level of information is given by the administrative border line. To adjust the permeability of the border line to pixels that are not border crossing points, we introduced the parameter $\delta_R \in [0,1]$. The value $C_i$ of a pixel $i$ on conductance map $C$ according to $R$ and $\delta_R$ is given by:
\begin{equation}
\label{C}
C_i =
\begin{cases}
\delta_R  R_i & \text{if } i \text{ is not a border crossing point}\\
R_i & \text{otherwise}
\end{cases}
\end{equation}

\noindent\textbf{Livestock movement modeling.} The last step consisted in assigning a potential route between every pair of nodes of our livestock mobility network using the conductance maps described in the previous section. To do so, we conducted a connectivity analysis based on concepts from electronic circuit theory \cite{McRae2007} using Circuitscape software (v4) (\url{https://pypi.org/project/Circuitscape/}, last accessed 18/02/2020). This approach has been widely used in wildlife corridor design \cite{Brodie2015,Mateo2015}, movement ecology, \cite{Bishop2015,McClure2016} and epidemiology \cite{Tatem2012}. 

For each pair of locations, represented by two pixels on the conductance map, Circuitscape computes a map of the total movement resistance accumulated from the origin and destination based on the electronic circuit theory applied on the conductance map \cite{McRae2007}. This map informs us about the potential for each pixel to be crossed during a livestock movement from the market of origin to the market of destination. We then normalized the map by its highest pixel value. 

Then, we multiplied each normalized connectivity map by the ratio of animals concerned (i.e. number of animals moving from the origin to the destination divided by the total number of animals). We finally summed all the maps. We obtained a final map of the potential path for livestock, presented in the next section, where the highest values indicate the highest potential for livestock movements. 

\subsection*{Identification of high potential areas} 

In animal health programs, land-use planning, or management of conflicts between farmers and herders, it is essential to be able to prioritize intervention zones. To do so, we need to spatially aggregate the information contained in the maps of potential paths for livestock movements in order to identify high potential areas. In this study, we spatially aggregated the maps of potential paths at regional level for Senegal, Gambia and Mauritania, using data downloaded from the GADM web platform (\url{https://gadm.org/index.html}, last accessed 18/02/2020). We thus obtained a distribution of values informing us about the level of activity within each administrative unit based on the potential for each $500 \times 500$ m$^2$ pixel to be crossed during a livestock movement. To facilitate the interpretation, the level of activity has been normalized by its maximum value and used to rank the different administrative units. We can also compute the level of normalized activity in each administrative unit based on the information provided by the livestock mobility network to compare the different approaches. In this case the activity is based on the total number of animals transiting in the administrative unit (sum of the in- and out-strength of the nodes located in the administrative unit).

To compare the different methods (landscape connectivity or network approaches) or the results obtained for different seasons, the distance between distributions of normalized activities (i.e rankings) can be assessed with the Kendall's $\tau$ coefficient \cite{Puka2011}. A value close to 1 means that the administrative units are ordered in the same way, while a value close to 0 means that there is no concordance in the rankings.

\subsection*{Sensitivity analysis} 

There are two main sources of uncertainty in the mapping of potential paths for livestock movements: the parameters $\delta_W$ and $\delta_R$ used to combine the different layers and the weights used to model the land use conductance (Table \ref{Tab1}). We used as reference the parameter values $\delta_W=0.8$ and $\delta_R=0.1$. This means that the walking layer based on land use information accounts for 80\% of the road network importance and the border has a very low permeability (10\% of the conductance of the road/walking layer $R$). The reference for the land use weights are displayed in Table \ref{Tab1} according to the season. For both sources of uncertainty, we rely on the Kendall's $\tau$ coefficient to compare the ranking of administrative units obtained with the reference distribution of activity with the ones obtained with different parameters and land use weight values. The two sources of uncertainty have been evaluated independently. For the parameters $\delta_W$ and $\delta_R$, we generated 25 rankings obtained with different pairs of values ranging between 0 and 1 by step of 0.25. For the land use weights, we changed one-at-a-time the weight of the different land use types by adding or subtracting an amount $\Delta=0.05$ or $\Delta=0.1$ from the original value.

\section*{Results}

\subsection*{Mobility Network analysis}

\begin{figure*}[!ht]
	\centering
	\includegraphics[width=12cm]{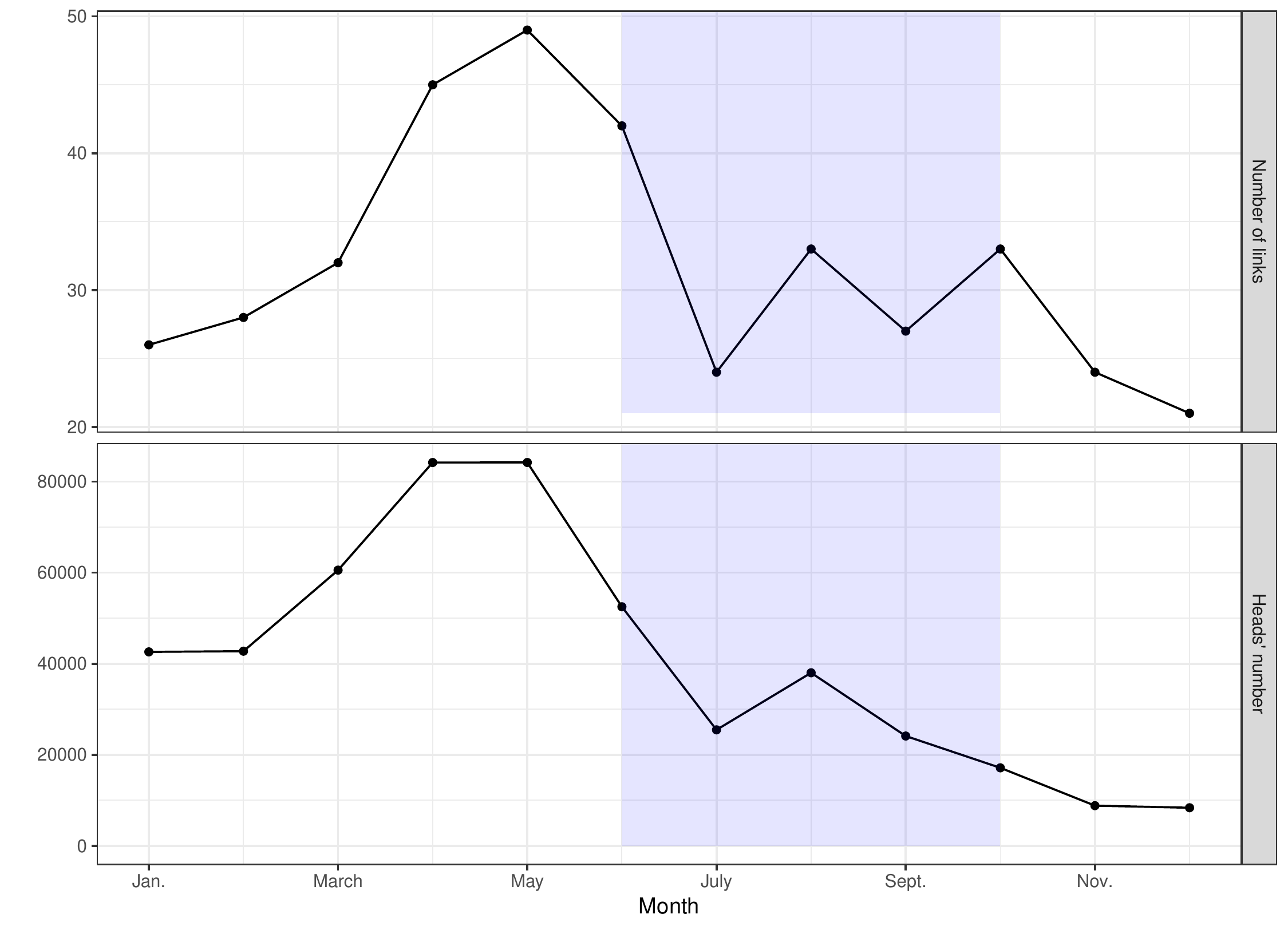} 
	\caption{\textbf{Network variation in 2014.} Number of links (top) and number of head displaced (bottom) depending on the month. The shaded area represents the wet season.
	\label{Fig4}}
\end{figure*}

Figure \ref{Fig4} shows the changes in the network measured throughout 2014, focusing on the number of links and animals transported each month. As can been seen, most of the activity is concentrated in the months before the wet season (April-June), when the scarcity of rainfall impedes the regeneration of pastures and animals are moved looking for better places. It is worth noting that the wet season (shaded area) is characterized by a dramatic reduction of links and animal movements. 

\begin{figure*}[!ht]
	\centering
	\includegraphics[width=15cm]{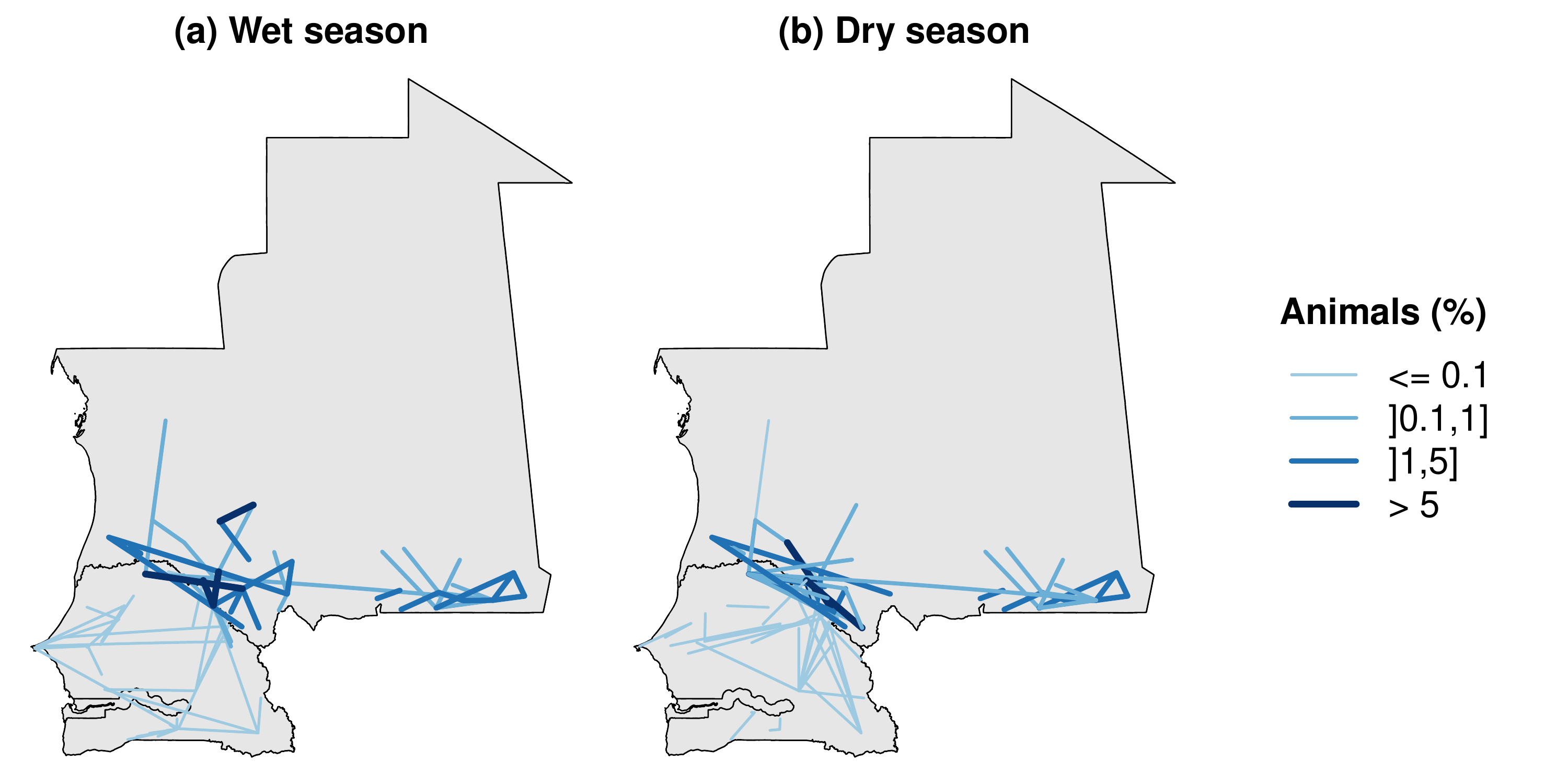} 
	\caption{\textbf{Cattle Mobility Networks in the wet (a) and dry (b) seasons.} The width and the color of a link is proportional to the number of animals displaced. The number of animals displaced from one node to another has been normalized by the total number of animals displaced and is expressed as a percentage. \label{Fig5}}
\end{figure*}

\begin{figure*}[!ht]
	\centering
	\includegraphics[width=14cm]{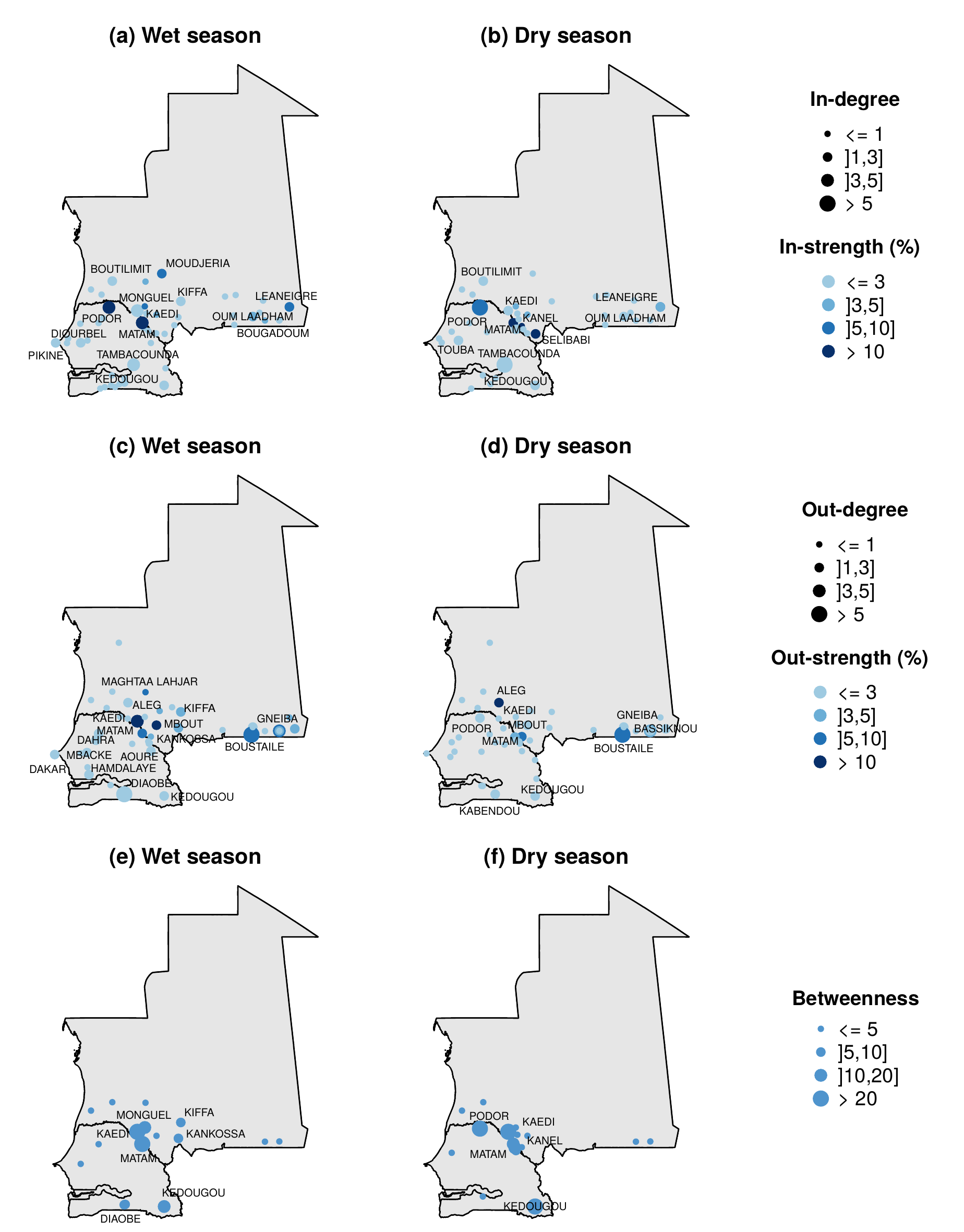} 
	\caption{\textbf{Node centrality analysis.} For each node, five centrality indices are displayed for the wet season ((a), (c) and (e)) and the dry season ((b), (d) and (f)): in-degree and in-strength (a)-(b), out-degree and out-stength (c)-(d) and the betweenness (e)-(f). Size of the dots is proportional to the degree (a)-(d) or the betweenness (e)-(f). Color of the dots corresponds to the in- and out-strength (a)-(d). In- and out-strength has been normalized by the total number of animal and are expressed in percentage. \label{Fig6}}
\end{figure*}

\begin{table}[!h]
	\begin{center}
		\begin{tabular}{lcccc}
			\hline
			Season & Nodes & Links & Volume & Top 10 links volume (\%)\\
			\hline
			All & 108 & 116 & 0.49 & 66.83\\
			Wet & 85 & 81 & 0.16 & 65.99\\
			Dry & 84 & 78 & 0.33 & 74.54\\
			\hline
		\end{tabular}
	\end{center}
	\caption{\textbf{Total number of nodes, links and volume of animals according to the season.} Each node represents an origin or a destination in the livestock mobility network. A link is created between two nodes if at least one animal moves from one node to another. The volume is expressed in million of head.}
	\label{Tab2}
\end{table}

Table \ref{Tab2} shows the total number of nodes, links and volume of animals displaced depending on the time period. We observe a similar number of links and nodes in the two seasonal networks. We observe however more than twice as many animals are displaced in dry season compared to the wet season. A visual representation of the network in the two seasons is shown in Figure \ref{Fig5}, where link colors and thickness correspond to the number of animals displaced (expressed as a percentage of the total). In both cases, the majority of the links corresponds to movements of small herds and accounted for less than 1$\%$ of the total volume. The top 10 links accounted for about 66$\%$ of the total volume of animals in the wet season and 75$\%$ in the dry season (Table \ref{Tab2}). The majority of the animal movements takes place in two areas. The first area is located around the Senegalese-Mauritanian border, with high trade activity between large cities in Mauritania (Nbeika, Boutilimit, Aleg, Mbout, Kaedi and Selibabi) and Senegal (Podor, Matam and Kanel). A major share of these movements involves transboundary movements between Podor and Mbout or between Matam and Kaedi and Mongel, for example. This observation applies to both seasons, but transboundary activity seems to be greater in the wet season than in the dry season. The second area showing major activity is located in southeastern Mauritania close to the border with Mali, involving cities such as Boustaile and Gneiba. It should also be noted that, although more moderate, there is also trade activity between Senegalese cities furthest from the border, such as Kedougou, Diaobe, Tambacounda for the South and Dakar, Diourbel and Touba for the West. That activity is more pronounced during the wet than the dry season.

\begin{figure*}[!ht]
	\centering
	\includegraphics[width=14cm]{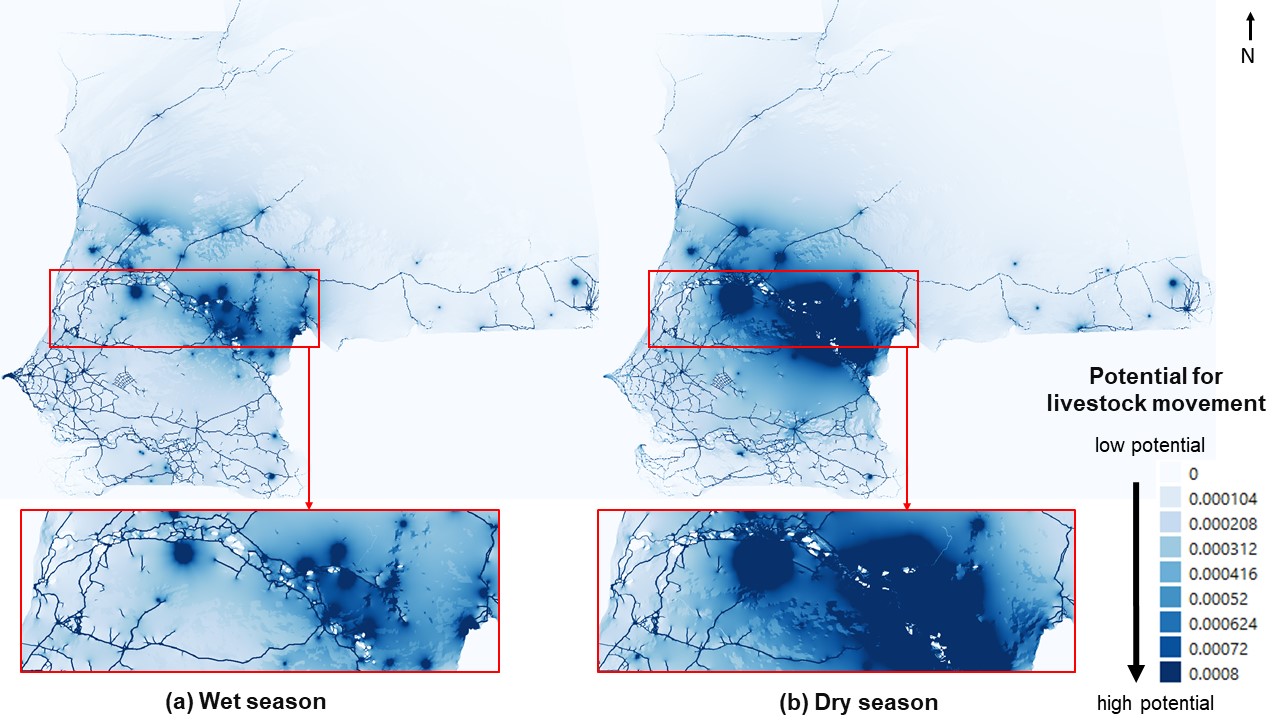} 
	\caption{\textbf{Maps of the potential paths for livestock movements according to the season.} (a) Wet season. (b) Dry season. The maps are based on the parameter values $\delta_W=0.8$ and $\delta_R=0.1$ and the land use weights presented in Table \ref{Tab1}. \label{Fig7}}
\end{figure*}

The role played by the different locations slightly changes from one season to another. Figure \ref{Fig6} shows the different locations highlighted according to their centrality. Most of the locations maintain their \enquote{activity} between the two seasons. This is particularly true for the largest market areas of Podor, Kaedi and Matam, located on the border between Senegal and Mauritania, but also for Kedougou in southern Senegal and Boustaile on the border between Mauritania and Mali. They represent major destinations for animal movements. It can be seen in Figure \ref{Fig6}e and \ref{Fig6}f that Podor, on the Senegalese/Mauritanian border is an important transit point during the dry season, but not during the wet season. This network analysis provides useful information about the livestock mobility network in Senegal and Mauritania. However, it does not enable explicit mapping of livestock movements.

\subsection*{Mapping potential paths for livestock movements}

We plotted in Figure \ref{Fig7} the maps of potential paths for livestock movements in the wet and dry seasons obtained with the landscape connectivity approach. The two maps show different potential movement patterns. For example, the area on the Senegal-Mauritania eastern border is less permeable in the wet season than in the dry season. Moreover, the wet season map shows more complex patterns of passage potential in that area. This was due to the presence of crop plots (see Figure \ref{Fig3}), or floodplains, that animals have to avoid during that season. This highlights the importance of the explicit mapping of network links according to landscape conductance, in order to spatially translate connectivity. For both seasons, the highest potential passages is located around the roads. This is even more pronounced for the wet season, during which some areas could not be crossed and animals are forced to use tracks alongside the roads. Whatever the season, the two maps show one large core area with high crossing potential located on the eastern side of the border between Senegal and Mauritania. Areas located in southern Senegal (Kedougou) and in the southeastern Mauritania (Boustaile) show a low  passage potential, while they clearly appear as central nodes in the livestock mobility network (Figure \ref{Fig6}). On the other hand, certain areas located around the Podor-Kaedi-Matam axis exhibit a high passage potential, yet it does not contain any origin or destination nodes. It is typically an area where animals pass through and crossbred, which our methodology enables us identify and delimit. This shows the relevance of landscape connectivity based approaches for identifying areas with a high potential for livestock movements.

\begin{figure*}[!ht]
	\centering
	\includegraphics[width=12cm]{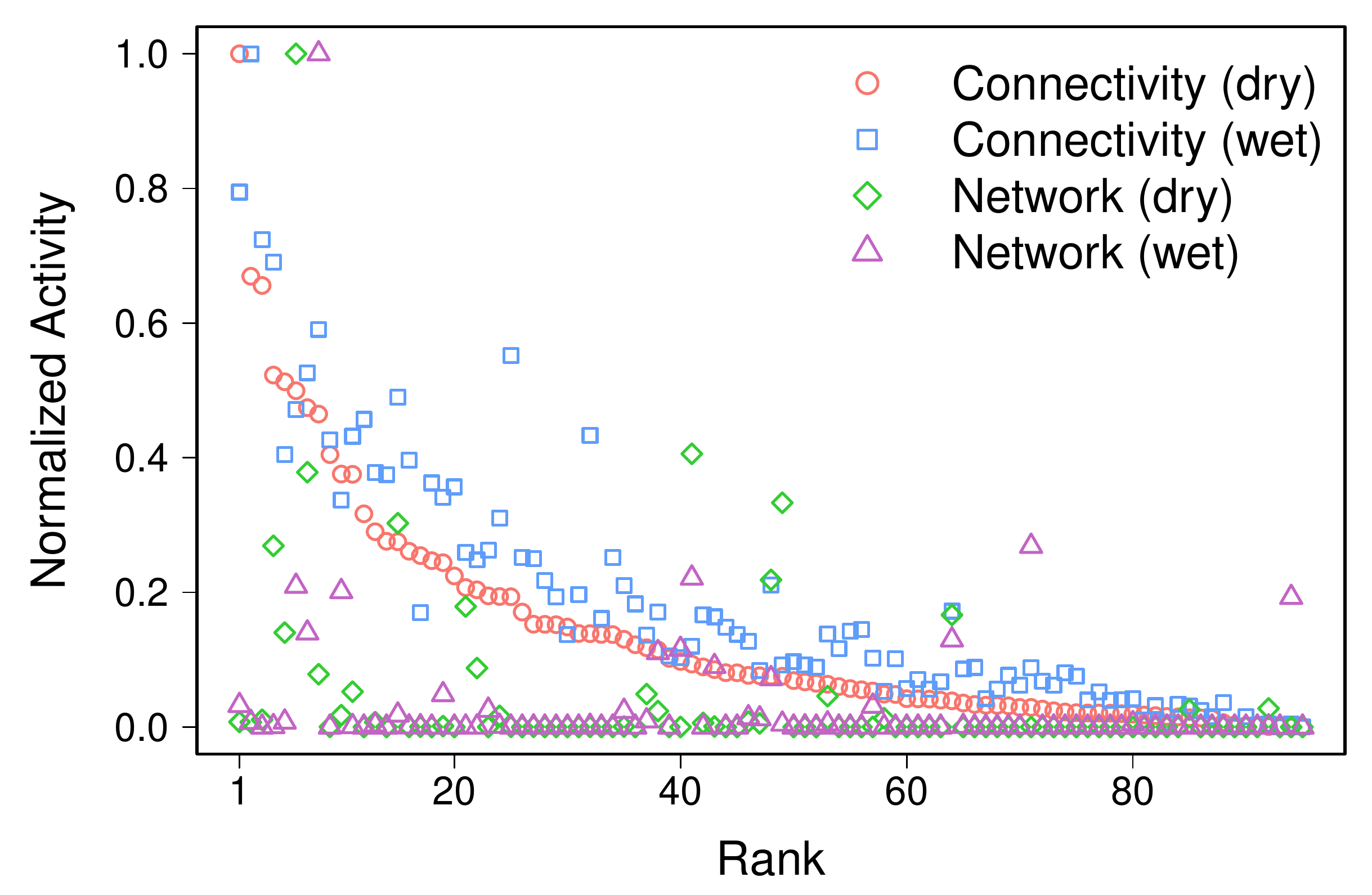} 
	\caption{\textbf{Rank-size distribution of the normalized activity obtained with the landscape connectivity and the network approaches.} The total activity (potential livestock movements for the landscape connectivity and total of out- and in-strength for the network approach) contained in each administrative unit have been considered and each distribution have been normalized by its maximum value. The values are ordered according to the activity obtained with the landscape connectivity approach in dry season. Values obtained with the landscape connectivity approach have been calculated with the parameter values $\delta_W=0.8$ and $\delta_R=0.1$ and the land use weights presented in Table \ref{Tab1}. \label{Fig8}}
\end{figure*}

\begin{figure*}[!ht]
	\centering
	\includegraphics[width=13cm]{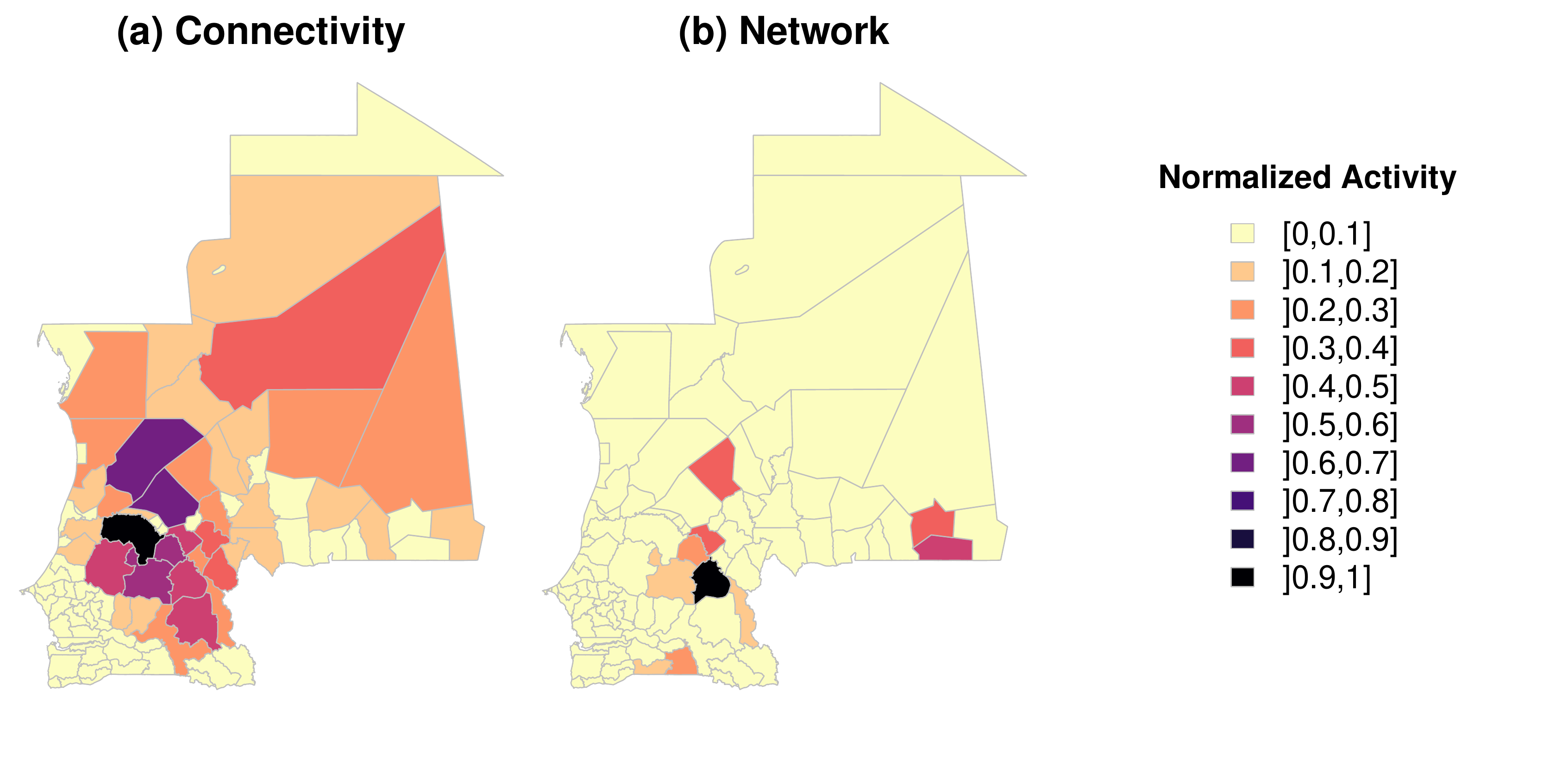} 
	\caption{\textbf{Maps of the normalized activity obtained with our method (a) and a network approach (b) in the dry season.} The total activity (potential livestock movements for the landscape connectivity and total of out- and in-strength for the network approach) contained in each administrative unit have been considered and each distribution have been normalized by its maximum value. Values obtained with the landscape connectivity approach have been calculated with the parameter values $\delta_W=0.8$ and $\delta_R=0.1$ and the land use weights presented in Table \ref{Tab1}. \label{Fig9}}
\end{figure*}

\subsection*{Identification of high potential areas} 

We plot in Figure \ref{Fig8} the rankings of regional administrative units obtained with the different methods (landscape connectivity and network approaches) in the dry and wet seasons. We observe that there was a large difference between administrative unit rankings obtained with the landscape connectivity and network approaches, whatever the season. This is not really surprising, since the two types of activity are not based on the same information, but it highlights the importance of spatially mapping potential paths to identify active areas in terms of animal movements. In particular, there are several units with no activity according to the mobility network that are in the top 10 for the activity measured with the landscape connectivity approach. Maps of the spatial distribution of activity measured with the two approaches in the dry season can be found in Figure \ref{Fig9}. To quantify these differences more rigorously, we computed the correlation between the different rankings with the Kendall's $\tau$ coefficient as described in the Material and methods section. Table \ref{Tab3} shows the correlation matrix comparing the four distributions displayed in Figure \ref{Fig8}. We observe a low correlation between connectivity and network approaches whatever the season, thus confirming the results observed in Figure \ref{Fig8}. We also note a strong correlation ($\tau=0.84$) between the rankings obtained with the landscape connectivity approach in wet and dry seasons. It is interesting to note that this correlation falls to 0.66 when comparing the network approach in the wet and dry seasons.  

\begin{table*}[ht]
	\caption{\textbf{Kendall rank correlation coefficient matrix.} Kendall's $\tau$ coefficient between the four rankings displayed in Figure \ref{Fig8} (Landscape connectivity approach and network approach in dry and wet seasons).Values in bracket correspond to the confidence interval of the correlation coefficient at 95\%.}
	\begin{center}
		\begin{tabular}{ccccc}
			& Connectivity (dry) & Connectivity (wet) & Network (dry) & Network (wet)\\ 
			Connectivity (dry) & 1 & 0.84 [0.76,0.89] & 0.4 [0.22,0.56] & 0.29 [0.09,0.46]\\ 
			Connectivity (wet) &  & 1 & 0.41 [0.23,0.57] & 0.31 [0.12,0.48]\\
			Network (dry) &  &  & 1 & 0.66 [0.53,0.76]\\ 
		\end{tabular}
	\end{center}
	\label{Tab3}
\end{table*}

\begin{figure*}[!ht]
	\centering
	\includegraphics[width=15cm]{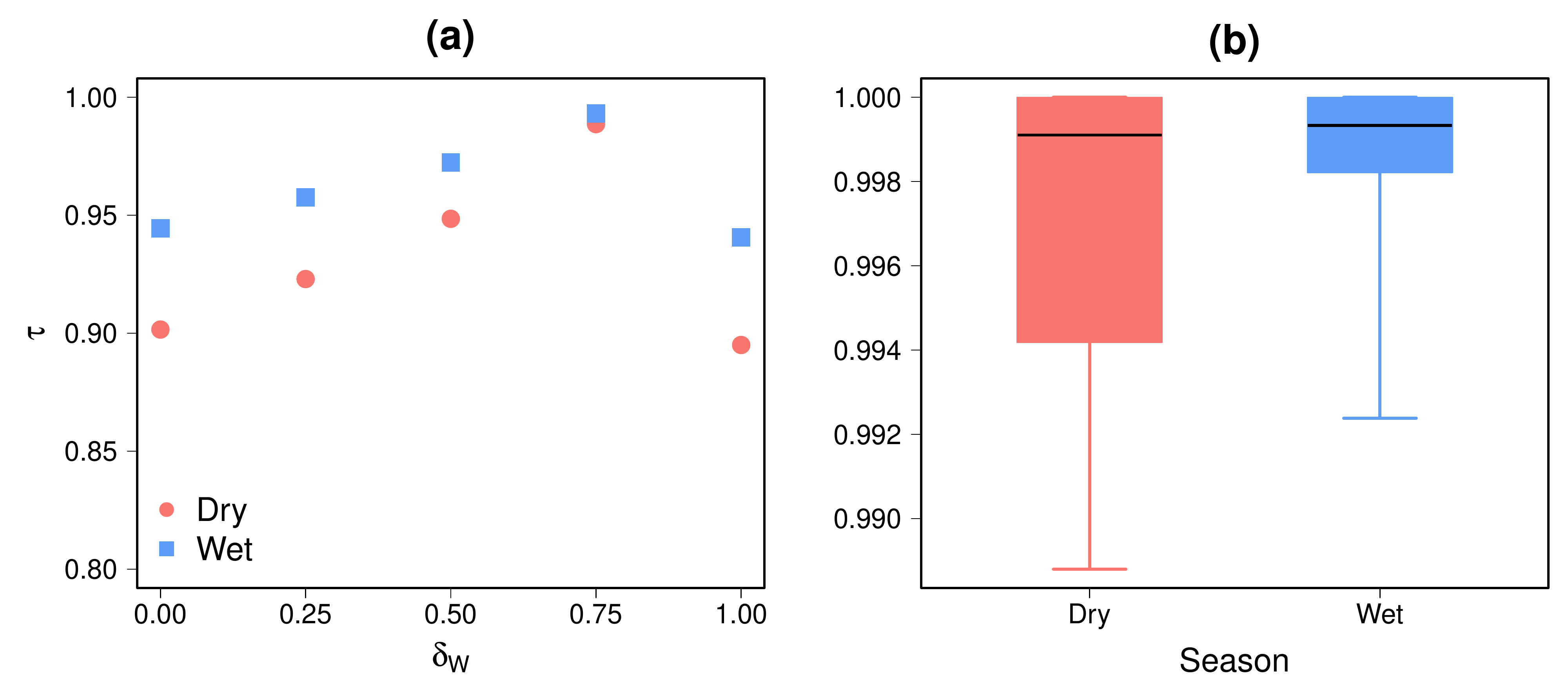} 
	\caption{\textbf{Parameters (a) and land use weights (b) sensitivity analysis in dry and wet seasons.} (a) Kendall's $\tau$ coefficient between the reference ranking and the ranking obtained with different parameter values as a function of $\delta_W$. For each $\delta_w$, the $\tau$ values have been averaged over $\delta_R$ values. The value of $\tau$ for each couple of ($\delta_W$,$\delta_R$) are available in Appendix (Tables S1 and S2). (b) Boxplots of the Kendall's $\tau$ coefficient between the reference ranking and the ranking obtained with different land use weight values.  Each boxplot is composed of the minimum, the lower hinge, the median, the upper hinge and the maximum. The value of $\tau$ for each land use weight values are available in Appendix (Tables S3 and S4). \label{Fig10}}
\end{figure*}

\subsection*{Results of the sensitivity analysis} 

Figure \ref{Fig10} shows the results of the parameters and land use weights sensitivity analysis in the dry and wet seasons. We observe in Figure \ref{Fig10}a that the similarity between the ranking of reference and the ones obtained with different $\delta_W$ values is globally high with a Kendall's $\tau$ coefficient ranging from 0.8 to 1. The similarity decreases slowly when $\delta_W$ deceases below the reference value, we observe a break of this trend when $\delta_W=1$. In this particular case, the results are no longer driven by the road network, leading to a modification in the potential movement patterns on a global scale. Note that since $\delta_R$ has almost no impact on the activity at a global scale (see Table S1 and S2 in Appendix for more details), for each $\delta_w$ value, the $\tau$ values have been averaged over $\delta_R$. It is however important to keep in mind that the effect of $\delta_R$ is probably higher at a local scale since it only affected areas close to the Mauritanian-Senegalese border. As can be observed in Figure \ref{Fig10}b changes in land use weight values have very little impact on the rankings (see Table S3 and S4 in Appendix for more details). In both cases, the sensitivity of the results to variations in parameters and land use weight values is higher in the dry than in the  wet seasons.

\section*{Discussion}

The precise description of livestock movement patterns has a central role in many applied questions. This is particularly true in Sahelian semi-arid regions, where it has become a crucial requirement to help decision-makers in dealing with conflicts between herders and farmers, or regarding the spread of animal diseases. The originality of the approach proposed in this article lies in the fine mapping of animal flows by weighting a conductance map by the number of head of livestock. The resulting raster map reflects the potential for livestock movement in each pixel according to its landscape connectivity and its position relative to the livestock mobility network. We illustrated our approach with a livestock mobility network in Senegal and Mauritania in the 2014 dry and wet seasons, which we combined with different land-use information (land cover, roads and borders). Our results demonstrate the robustness of our approach in identifying and ranking areas according to their potential for livestock movement.  Other applications from our methodology are now possible. For example, we could cross the information contained in our potential maps with risk factors for the spread of diseases like Rift Valley fever \cite{Tran2016}. It will conduct to identify areas with the highest risk of disease transmission. When crossing the maps stemming from the landscape connectivity approach with maps of cropped areas, we can also identify priority zones where passage corridors have to be settled and secured, as these zones have the highest risk of conflicts between farmers and breeders.

\subsection*{Limitations of the study}

It needs to be kept in mind that our approach is highly dependent on the data being used and their resolution. The resolution of the conductance map, at 500 meters in our study, depends on the resolution of the land cover map and might not enable the consideration of very fine paths. Our results showed that the potential map was mostly driven by the road network, which can also be a major source of uncertainty. 

Many factors drive mobility dynamics: landscape configuration, road quality, need for food, need for watering points, border crossing, religious feasts, etc. The conductance map has to include all these mobility-driven factors. For this study, we were able to collect most of the geographical layers for each of these factors, except that of the watering points (boreholes and ponds). Consequently, the maps obtained in this study do not take into consideration the need to pass through watering points, especially during the dry season. This is an important drawback counterbalanced by the fact that Senegal and Mauritania have a very dense grid of boreholes.

Another difficulty is the reliability of the mobility data. Mobility data were collected using two different approaches in Senegal and Mauritania. For the Mauritanian case, a synthetic survey was conducted  by the National Livestock and Veterinary Research Centre (CNERV) and compared with health certificates collected by Veterinarian Offices. In the case of Senegal, paper copies of sanitary movement permits (LPS) were collected by ad-hoc activities. These certificates provided information about origins and destinations, and we do not know if the composition of the herd changed during the journey due to animal sales. Furthermore, there was no proof that the herds actually reached their destination. Another bias in the data was linked to the fact that this data set did not include undeclared movements (for herds that did not have a sanitary movement permit).

Lastly, construction of the conductance map, which is the basis of the proposed methodology, relies on resistance weights given by experts. It should be noted that the main purpose of this article was to propose a methodology and we did not try to increase the number of experts. Nevertheless, we showed that small variations applied one at a time to the land use weight values have no significant effect on the rankings. To use the presented method for operational purposes, concerted thought needs to be given to the weights to be assigned, and a multivariate sensitivity analysis of these weights needs to be integrated into the approach.

\subsection*{Concluding remarks}

The identification of high potential for livestock movements is a core issue for decision-makers, whether in the field of animal health or territorial planning. Our approach opens up some interesting perspectives for modeling potential animal passage in semi-arid regions experiencing a lack of specific data on livestock movements. It is, however, important to note that a large share of livestock remains in its zone of origin. These sedentary animals are often in contact with transhumant animals that cross their territory. This information should be added, to complete the map of the potential for livestock movements provided in this study.

\section*{Acknowledgements}

ML thanks the French National Research Agency for its financial support (project NetCost, ANR-17-CE03-0003 grant). IS thanks the Veterinary Services in Senegal for the financial support (Government of Senegal Budget). The work of AA is partially funded by the EU grant H2020-727393 PALE-Blu. This project has received funding from the European Union’s Horizon 2020 research and innovation programme under grant agreement MOOD No 874850. The contents of this publication are the sole responsibility of the authors and do not necessarily reflect the views of the European Commission. A special thank goes to Peter Biggins for correcting English.

\section*{Data availability}

Code and data are available at \url{www.maximelenormand.com/Codes}

\bibliographystyle{unsrt}
\bibliography{MPP}

\newpage
\onecolumngrid

\makeatletter
\renewcommand{\fnum@figure}{\sf\textbf{\figurename~\textbf{S}\textbf{\thefigure}}}
\renewcommand{\fnum@table}{\sf\textbf{\tablename~\textbf{S}\textbf{\thetable}}}
\makeatother

\setcounter{figure}{0}
\setcounter{table}{0}
\setcounter{equation}{0}

\section*{Appendix}

\subsection*{Supplementary figures}

\begin{figure}[!ht]
	\centering
	\includegraphics[width=12cm]{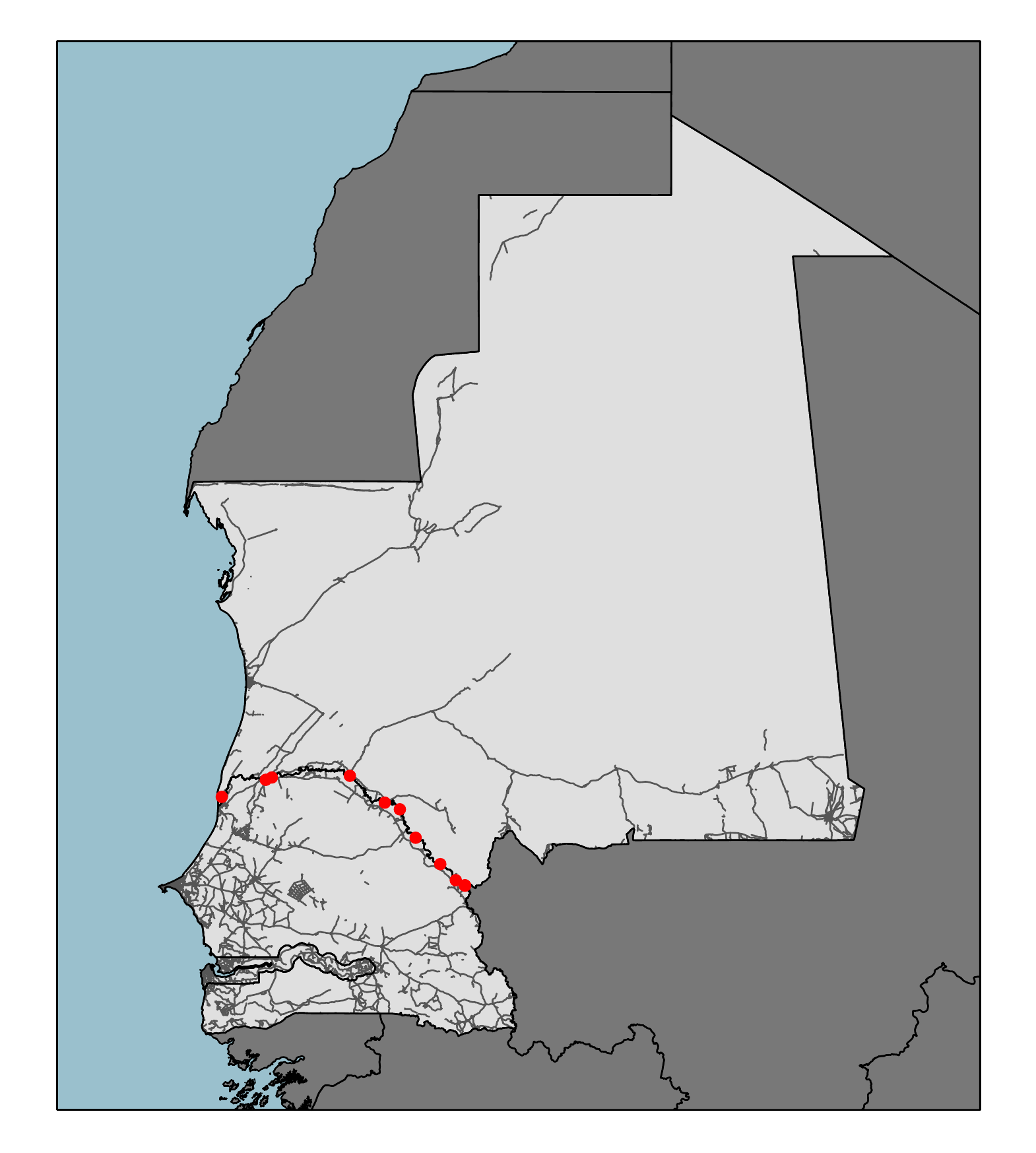} 
	\caption{\textbf{Senegal, Gambia and Mauritania's main road network.} The red points represent the border checkpoints. The main road network in Senegal, Gambia and Mauritania has been downloaded from OpenStreetMap (data available online at \url{https://www.openstreetmap.org}, last accessed 18/02/2020). OpenStreetMap is made available under the Open Database License \url{http://opendatacommons.org/licenses/odbl/1.0/}. Any rights in individual contents of the database are licensed under the Database Contents License: \url{http://opendatacommons.org/licenses/dbcl/1.0/}. \label{FigS1}}
\end{figure}

\newpage
\subsection*{Supplementary Tables}

\begin{table}[!h]
	\begin{center}
		\begin{tabular}{lccccc}
			\hline
			& $\delta_R=0$ & $\delta_R=0.25$ & $\delta_R=0.5$ & $\delta_R=0.75$ & $\delta_R=1$\\
			\hline
			
			$\delta_W=0$ & 0.9015 & 0.9015 & 0.9015 & 0.9028 & 0.9006 \\ 
			$\delta_W=0.25$ & 0.9230 & 0.9230 & 0.9230 & 0.9230 & 0.9230 \\
			$\delta_W=0.5$ & 0.9489 & 0.9489 & 0.9489 & 0.9485 & 0.9471 \\
			$\delta_W=0.75$ & 0.9892 & 0.9892 & 0.9892 & 0.9897 & 0.9857 \\
			$\delta_W=1$ & 0.8952 & 0.8952 & 0.8952 & 0.8952 & 0.8943 \\ 
			
			\hline
		\end{tabular}
	\end{center}
	\caption{\textbf{Parameter sensitivity analysis in dry season.} Kendall's $\tau$ coefficient between the ranking obtained with the reference distribution ($\delta_W=0.8$ and $\delta_R=0.1$) and the rankings obtained with different couples ($\delta_W$,$\delta_R$) values ranging between 0 and 1 in dry season. In all cases the conductance maps have been computed with the land use weights in dry season as defined in Table 1 in the main text.}
	\label{TabS1}
\end{table}

\begin{table}[!h]
	\begin{center}
		\begin{tabular}{lccccc}
			\hline
			& $\delta_R=0$ & $\delta_R=0.25$ & $\delta_R=0.5$ & $\delta_R=0.75$ & $\delta_R=1$\\
			\hline
			
			$\delta_W=0$ & 0.9449 & 0.9449 & 0.9449 & 0.9454 & 0.9422 \\
			$\delta_W=0.25$ & 0.9579 & 0.9579 & 0.9579 & 0.9579 & 0.9566 \\ 
			$\delta_W=0.5$ & 0.9727 & 0.9727 & 0.9727 & 0.9727 & 0.9709 \\
			$\delta_W=0.75$ & 0.9942 & 0.9942 & 0.9937 & 0.9937 & 0.9901 \\
			$\delta_W=1$ & 0.9409 & 0.9409 & 0.9404 & 0.9409 & 0.9404 \\ 
			
			\hline
		\end{tabular}
	\end{center}
	\caption{\textbf{Parameter sensitivity analysis in wet season.} Kendall's $\tau$ coefficient between the ranking obtained with the reference distribution ($\delta_W=0.8$ and $\delta_R=0.1$) and the rankings obtained with different couples ($\delta_W$,$\delta_R$) values ranging between 0 and 1 in wet season. In all cases the conductance maps have been computed with the land use weights in wet season as defined in Table 1 in the main text.}
	\label{TabS2}
\end{table}

\begin{table}[!h]
	\begin{center}
		\begin{tabular}{lcccc}
			\hline
			Land use type & $\Delta=-0.1$ & $\Delta=-0.05$ & $\Delta=0.05$ & $\Delta=1$\\ 
			\hline
			
			Coastal strip & 1 & 1 & 1 & 1 \\ 
			Mangrove & 1 & 1 & 0.9996 & 0.9991 \\ 
			Water bodies & 0.9991 & 0.9996 & 0.9996 & 0.9991 \\ 
			Irrigated croplands & - & - & - & - \\ 
			Croplands & 0.9892 & 0.9933 & - & - \\ 
			Forest area & 0.9987 & 0.9991 & 0.9987 & 0.9973 \\ 
			Mosaic croplands \& grassland & 0.9888 & 0.9937 & - & - \\ 
			Open grassland & 0.9892 & 0.9928 & - & - \\ 
			Dune and peneplain pastures & 0.9928 & 0.9955 & 0.9942 & 0.9892 \\ 
			Dune and gravel pastures & 0.9897 & 0.9946 & - & - \\ 
			Salt land & 0.9996 & 1 & - & - \\ 
			Bare rock & 0.9982 & 0.9991 & 0.9987 & 0.9982 \\ 
			Urban area & 1 & 1 & 1 & 1 \\ 
			Major rivers & - & - & - & - \\ 
			
			\hline
		\end{tabular}
	\end{center}
	\caption{\textbf{Land use weight sensitivity analysis in dry season.} Kendall's $\tau$ coefficient between the ranking obtained with the reference distribution (land use weight in dry season as defined Table 1 in the main text) and the rankings obtained with small variation $\Delta$ applied on the original values ranging between -0.1 and 0.1 when applicable. In all cases the maps of potential paths have been obtained with the parameters $\delta_W=0.8$ and $\delta_R=0.1$.}
	\label{TabS3}
\end{table}

\begin{table}[!h]
	\begin{center}
		\begin{tabular}{lcccc}
			\hline
			Land use type & $\Delta=-0.1$ & $\Delta=-0.05$ & $\Delta=0.05$ & $\Delta=1$\\ 
			\hline
			
			Coastal strip & 0.9996 & 0.9996 & 1 & 1 \\ 
			Mangrove & 1 & 1 & 1 & 1 \\ 
			Water bodies & 1 & 1 & 1 & 1 \\ 
			Irrigated croplands & - & - & - & - \\ 
			Croplands & 0.9987 & 0.9996 & 0.9982 & 0.9955 \\ 
			Forest area & 0.9982 & 0.9987 & 0.9991 & 0.9987 \\ 
			Mosaic croplands \& grassland & 0.9996 & 0.9996 & 0.9978 & 0.9964 \\ 
			Open grassland & 0.9937 & 0.9960 & - & - \\ 
			Dune and peneplain pastures & 0.9969 & 0.9987 & 0.9969 & 0.9924 \\ 
			Dune and gravel pastures & 0.9942 & 0.9969 & 0.9982 & 0.9951 \\ 
			Salt land & 0.9991 & 0.9996 & 1 & 0.9996 \\ 
			Bare rock & 0.9982 & 0.9991 & 0.9996 & 0.9987 \\ 
			Urban area & 1 & 1 & 1 & 1 \\ 
			Major rivers & - & - & - & - \\ 
			
			\hline
		\end{tabular}
	\end{center}
	\caption{\textbf{Land use weight sensitivity analysis in wet season.} Kendall's $\tau$ coefficient between the ranking obtained with the reference distribution (land use weight in wet season as defined Table 1 in the main text) and the rankings obtained with small variation $\Delta$ applied on the original values ranging between -0.1 and 0.1 when applicable. In all cases the maps of potential paths have been obtained with the parameters $\delta_W=0.8$ and $\delta_R=0.1$.}
	\label{TabS4}
\end{table}

\end{document}